\documentstyle[manuscript,aps,prc,tighten,amssymb,graphicx]{revtex}
\begin{document}
\def\eps{\epsilon}
\def\epsnn{\epsilon_{NN}}
\def\la{\Lambda}
\def\si{\Sigma}
\def\sim{{\Sigma^-}}

\draft

\title{Hyperon stars in the Brueckner-Bethe-Goldstone theory}

\author{M. Baldo and G. F. Burgio}
\address{Istituto Nazionale di Fisica Nucleare, Sezione di Catania and 
Dipartimento di Fisica, \\ 
Universit\`a di Catania, Corso Italia 57, I-95129 Catania, Italy}
\author{H.-J. Schulze }
\address{Departament d'Estructura i Constituents de la Mat\`eria,
Universitat de Barcelona, \\
Av. Diagonal 647, E-08028 Barcelona, Spain}

\date{\today}

\maketitle

\begin{abstract}
  In the framework of the Brueckner-Bethe-Goldstone theory, we 
  determine a fully microscopic equation of state for asymmetric and
  $\beta$-stable nuclear matter containing $\sim$ and $\la$ hyperons. 
  We use the Paris and the new Argonne $Av_{18}$ 
  two-body nucleon interaction, whereas the nucleon-hyperon interaction
  is described by the Njimegen soft-core model. 
  We stress the role played by the three-body nucleon interaction, 
  which produces a strong repulsion at high densities. 
  This enhances enormously the hyperon population, and produces a strong 
  softening of the equation of state, which turns out almost independent 
  on the nucleon-nucleon interaction. 
  We use the new equation of state in order to calculate
  the structure of static neutron stars.  
  We obtain a maximum mass configuration with $M_{\rm max}$ = 1.26 (1.22) 
  when the Paris ($Av_{18}$) nucleon potential is adopted. 
  Central densities are about 10 times normal nuclear matter density. 
  Stellar rotations, treated within a perturbative approach, increase 
  the value of the limiting mass by about $12 \%$. 
\end{abstract}

\bigskip
\bigskip
\bigskip


\section{Introduction}

The nuclear matter equation of state (EOS) is the fundamental input for 
building models of neutron stars (NS) \cite{shapiro}. 
These compact objects, among the densest ones in the universe, are produced 
during the gravitational collapse of massive stars, which explode into 
supernovae at the end of their evolution. 
Neutron stars are observed as pulsars: because of their fast 
rotation they emit only in particular directions regularly spaced pulses of
electromagnetic radiation. 
Although almost 700 pulsars have been detected
so far, their gravitational mass can be inferred only from observation of 
a few binary systems \cite{masses}. 
The observed NS masses are typically $\approx (1-2) M_{\odot}$
(where $M_{\odot}$ is the mass of the sun, $M_\odot = 1.99\times 10^{33}$g).
Above $3 M_\odot$, NS are commonly believed to collapse into black holes.
Typical radii of NS are thought to be of order 10 km, although direct 
measurements do not exist, whereas the central density is a few times 
normal nuclear matter density ($\rho_0\approx 0.17\;{\rm fm}^{-3}$). 
This requires a detailed knowledge of the EOS for densities $\rho \gg \rho_0$.
 
This is a very hard task from the theoretical point of view. 
For instance, the present uncertainty on 
the equation of state at high density implies an uncertainty 
on the value of the maximum mass, important for distinguishing between 
neutron star or a black hole formation. 
In fact, whereas at densities $\rho \approx \rho_0$ the matter consists 
mainly of nucleons and leptons, at higher densities several species 
of particles may appear due to the fast rise of the baryon chemical potentials 
with density.
Among these new particles are strange baryons, namely, the $\la$, $\si$ and
$\Xi$ hyperons. 
Due to its negative charge, the $\sim$ hyperon is the 
first strange baryon expected to appear with increasing density in the 
reaction $n+n \rightarrow p+\sim$,
in spite of its substantially larger mass compared to the neutral $\la$ 
hyperon ($M_\sim=1197\;{\rm MeV}, M_\la=1116\;{\rm MeV}$).
Other species in stellar matter may appear,
like $\Delta$ isobars along with pion and kaon condensations.
Moreover, at very high densities, nuclear matter is expected to undergo
a transition to a quark-gluon plasma \cite{quark}. 
However, the exact value of the transition density is still unknown 
because of some
technical problems in the QCD lattice calculations for finite baryon density. 
In this paper we disregard these phenomena, because they lie outside
the scope of Brueckner theory that is applied here. 
In particular, we concentrate our investigation on the production 
of strange baryons. 
We assume that a baryonic description of nuclear matter 
holds up to densities as those encountered in the core of neutron stars. 

In a previous article \cite{bbs} we presented a microscopic investigation 
within an extended Brueckner-Hartree-Fock (BHF) scheme for
determining the chemical potentials of the different
baryons ($n,p,\sim,\la$) in a fully self-consistent manner.
We used in our calculations the Paris \cite{lac80} and the Argonne
$v_{14}$ \cite{wir84} nucleon-nucleon (NN) interaction, modified 
by three-body forces (TBF), 
in order to get the correct saturation point of nuclear matter \cite{threeb}. 
In that paper \cite{bbs} we adopted the Nijmegen soft-core \cite{mae89} 
potentials for describing the nucleon-hyperon interaction, whereas
no hyperon-hyperon potential was taken into account, due to lack of 
restricting experimental data. 
We mainly concentrated on the calculation of the 
onset density of the $\sim$ and $\la$ hyperons. 
We found that thresholds are reached 
at densities beginning at about 2--3 times normal nuclear matter density,
for all the different nuclear equations of state considered.

In this paper we proceed further and present results concerning the 
equation of state of asymmetric and $\beta$-stable nuclear matter 
containing $\sim$ and $\la$ hyperons, obtained in the BHF theoretical scheme. 
We adopt the Paris and the new Argonne $v_{18}$ \cite{av18} NN potentials, 
eventually modified by nucleon TBF according to the Urbana model, 
and the Njimegen nucleon-hyperon potentials. 
In these calculations still no hyperon-hyperon interaction is included. 
We determine microscopically the chemical potentials of the various species. 
Their concentration is inferred by imposing the conditions of chemical 
equilibrium, along with charge neutrality and baryon number conservation. 
In general, we observe a softening of the equation of state with respect to the
pure nucleonic case because of the increased number of baryonic species. 
The main result of our work is that in the presence of hyperons the inclusion 
of the nucleon TBF does not produce any significant change in the 
equation of state with respect to the case with only two-body forces. 
This is quite astonishing because, in the pure nucleon case, the repulsive
character of TBF at high density increases the stiffness of the EOS,
thus changing dramatically the equation of state \cite{threeb}. 
However, when hyperons are included, the presence of TBF among nucleons 
enhances the population of $\sim$ and $\la$ because of the increased 
neutron and proton chemical potentials with respect to the
case without TBF, thus decreasing the nucleon population.
The net result is that the equation of state looks very similar to the case
without TBF, but the chemical composition of 
matter containing hyperons is very different when TBF are included.
In the latter case, the hyperon populations are larger than in the case 
with only two-body forces. 
This has very important consequences for the structure of the neutron stars. 
Of course, this scenario could partly change if 
hyperon-hyperon interactions were known
or if TBF would be included also for hyperons, but this is 
beyond our current knowledge of the strong interaction.   

We apply our new equations of state to the calculation of the static 
properties of neutron stars having hyperonic cores. 
The values of the 
maximum mass configuration, i.e., mass, radius and central density, 
turn out almost independent on the chosen two-body nucleon-nucleon potential.
On the other hand, the values of the central densities
are quite different when TBF are taken into account, producing less compact
stars than in the case without TBF. 
Therefore hyperon stars, i.e., nucleon stars with hyperonic cores, 
collapse earlier than pure nuclear stars. 
Stellar rotations, which are treated
here within a modified Hartle-Thorne method \cite{wegl}, increase the value
of the maximum mass by about $12 \%$. 
However only configurations rotating at 
their Kepler frequency have been calculated.

This paper is organized as follows. 
In Section \ref{s:forma} we review some formalism
in the BHF scheme with hyperons, with a discussion of the method chosen for 
the calculation of the baryon chemical potentials. 
We also discuss the equilibrium compositions among the different baryonic
species and the equation of state. 
In Section \ref{s:res} we illustrate our results. 
In particular, in paragraph \ref{s:spp}
we discuss the behavior of the single-particle potentials for all involved 
species as calculated in the Brueckner theory. 
The matter composition is illustrated in paragraph \ref{s:comp}, 
along with the equation of state. 
Special emphasis is put on discussing the role of the nucleonic TBF
in paragraph \ref{s:tbf}.  
In paragraph \ref{s:nstar} we analyze the static properties of neutron stars 
and their variations caused by the rotations. 
Our conclusions are finally drawn in Section \ref{s:conc}.

\section{Formalism}
\label{s:forma}

For a detailed account of the extended Brueckner theory including hyperons we 
refer the reader to Refs.~\cite{bbs} and \cite{hjo95}. 
Here we repeat only the basic formulae and give some details concerning
the particular application to neutron star physics.
It turns out that in this case, besides the nucleons, $N=n,p$, only the
$Y=\sim,\la$ hyperons appear as stable particles in the matter, 
limiting the baryonic Fermi seas to these four species.
(Other types of hyperons appear in virtual intermediate states, of course).
We remind also that for the present work no hyperon-hyperon potentials are
used, leading to simplifiction in some places.

Then, the extended Brueckner scheme requires as input the nonrelativistic
nucleon-nucleon and nucleon-hyperon potentials.
With these potentials, the various $G$ matrices are evaluated by solving
numerically the Bethe-Goldstone equation \cite{bruck}, written schematically
\begin{equation}
  G_{ab}[W] = V_{ab} + \sum_c \sum_{p,p'} 
  V_{ac} \Big|pp'\Big\rangle 
  { Q_c \over W - E_c +i\epsilon} 
  \Big\langle pp'\Big| G_{cb}[W] \:, 
\label{e:g}
\end{equation}
where the indices $a,b,c$ indicate pairs of baryons
and the angle-averaged Pauli operator $Q$ and energy $E$ 
determine the propagation of intermediate baryon pairs.
In a given nucleon-hyperon channels $c=(NY)$ one has, for example,
\begin{eqnarray}
  E_{(NY)} &=& m_N + m_Y + {k_N^2\over2m_N} + {k_Y^2\over 2m_Y} +
  U_N(k_N) + U_Y(k_Y) \:.
\label{e:e}
\end{eqnarray}
The hyperon single-particle potentials within the continuous choice
are given by
\begin{eqnarray}
  U_Y(k) &=& \sum_{N=n,p} U_Y^{(N)}(k) 
  = {\rm Re}\, \sum_{N=n,p}\sum_{k'<k_F^{(N)}} 
  \Big\langle k k' \Big| G_{(NY)(NY)}\left[E_{(NY)}(k,k')\right] 
  \Big| k k' \Big\rangle 
\label{e:uy}
\end{eqnarray}
and similar expressions of the form
\begin{eqnarray}
  U_N(k) &=& \sum_{N'=n,p} U_N^{(N')}(k) + \sum_{Y=\sim,\la} U_N^{(Y)}(k) 
\label{e:un}
\end{eqnarray}
apply to the nucleon single-particle potentials.
The nucleons feel therefore direct effects of the other nucleons as well as 
of the hyperons in the environment, whereas for the hyperons there are only 
nucleonic contributions, because of the missing hyperon-hyperon potentials.

These equations (\ref{e:g}--\ref{e:un}) define the BHF scheme with the 
continuous choice of the single-particle energies.  
Due to the occurrence of $U_N$ and $U_Y$ in Eq.~(\ref{e:e}) they constitute 
a coupled system that has to be solved in a self-consistent manner.
In our previous work \cite{bbs} those equations were solved for zero
hyperon fraction, since we were interested only in their onset density.
In the present paper we proceed further and perform calculations
for arbitrary nucleon and hyperon concentrations. 
Therefore the above equations must be solved for several 
Fermi momenta of the particles involved. 

Once the different single-particle potentials are known,
the total nonrelativistic baryonic energy density, $\epsilon$,  
and the total binding energy per baryon, 
$B/A$, can be evaluated: 
\begin{eqnarray}
 {B\over A} &=& {\eps\over \rho_n+\rho_p+\rho_\sim+\rho_\la} \:,
\\
 \eps &=& \sum_{i=n,p,\sim,\la} \int_0^{k_F^{(i)}}\!\! {dk\,k^2\over\pi^2} 
 \left( m_i + {k^2\over{2m_i}} + {1\over2}U_i(k) \right) 
 = \eps_{NN} + \eps_{NY} 
\end{eqnarray}
with
\begin{mathletters}
\begin{eqnarray}
 \eps_{NN} &=& \sum_{N=n,p} \int_0^{k_F^{(N)}}\!\! {dk\,k^2\over\pi^2}
 \left( m_N + {k^2\over{2m_N}} + {1\over2} 
 \left[ U_N^{(n)}(k) + U_N^{(p)}(k) \right] \right) \:,
\label{e:epsnn}
\\
 \eps_{NY} &=& 
 \sum_{Y=\sim,\la} \int_0^{k_F^{(Y)}}\!\! {dk\,k^2\over\pi^2}
 \left( m_Y + {k^2\over 2m_Y} \right) +
 \sum_{N=n,p} \int_0^{k_F^{(N)}}\!\! {dk\,k^2\over\pi^2} 
 \left[ U_N^{(\sim)}(k) +  U_N^{(\la)}(k) \right] 
\label{e:epsny1}
\\
 &=& 
 \sum_{Y=\sim,\la} \int_0^{k_F^{(Y)}}\!\! {dk\,k^2\over\pi^2}
 \left( m_Y + {k^2\over 2m_Y} + 
 \left[ U_Y^{(n)}(k) + U_Y^{(p)}(k) \right] \right) \:.
\label{e:epsny2}
\end{eqnarray}
\label{e:epsnny}
\end{mathletters}
Here we have split the energy density into a part due to the action
of nucleon-nucleon forces, $\eps_{NN}$,
and due to nucleon-hyperon forces, $\eps_{NY}$.
These quantities depend on the total baryon density of the system, 
$\rho=\rho_N+\rho_Y$, 
($\rho_N = \rho_n + \rho_p$, $\rho_Y = \rho_\sim + \rho_\la$), and on
the baryon fractions $x_i=\rho_i/\rho$, $i=p,\sim,\la$. 
However, due to the fact that the single-particle potentials
$U_N^{(n)}$ and $U_N^{(p)}$ depend only indirectly and therefore very weakly 
on the hyperon partial densities, the energy density $\eps_{NN}$ depends 
to a good approximation only on the nucleonic partial density and the 
proton fraction within the nucleonic subsystem: 
$\eps_{NN} = \eps_{NN}(\rho_N,x=\rho_p/\rho_N)$,
whereas the dependence on the hyperonic partial densities is concentrated
in $\eps_{NY}$.

This facilitates the determination of the chemical potentials.
They are given by 
\begin{mathletters}
\begin{eqnarray}
  \mu_n(\rho,x_p,x_\sim,x_\la) &=& {\partial \epsilon \over \partial \rho_n}
  \approx \mu_n(\rho_N,x) + U_n^{(\sim)}(k_F^{(n)})+U_n^{(\la)}(k_F^{(n)}) \:,
\label{e:mun}
\\
  \mu_p(\rho,x_p,x_\sim,x_\la) &=& {\partial \epsilon \over \partial \rho_p}
  \approx \mu_p(\rho_N,x) + U_p^{(\sim)}(k_F^{(p)})+U_p^{(\la)}(k_F^{(p)}) \:,
\label{e:mup}
\\
  \mu_Y(\rho,x_p,x_\sim,x_\la) &=& {\partial \epsilon \over \partial \rho_Y}
  \approx m_Y + {{k_F^{(Y)}}^2 \over 2m_Y} + 
  U_Y^{(n)}(k_F^{(Y)}) + U_Y^{(p)}(k_F^{(Y)})  \:,
\label{e:muy}
\end{eqnarray}
\end{mathletters}
where Eq.~(\ref{e:epsny1}) (\ref{e:epsny2}) was used in order to arrive at 
Eqs.~(\ref{e:mun},\ref{e:mup}) (\ref{e:muy}).

The contributions to the chemical potentials due to $\eps_{NN}$
are then the same as in the system without hyperons, namely
\begin{mathletters}
\begin{eqnarray}
  \mu_n(\rho_N,x) &=& {\partial \epsnn \over \partial \rho_n}
  = \left( 1 + \rho_N {\partial\over\partial\rho_N} 
  - x{\partial\over\partial x} \right) 
  \left.{B\over A}\right|_{\rho_Y=0} \:,
\\
  \mu_p(\rho_N,x) &=& {\partial \epsnn \over \partial \rho_p}
  = \left(1 + \rho_N{\partial\over\partial\rho_N} 
  + (1-x){\partial\over\partial x} \right) 
  \left.{B\over A}\right|_{\rho_Y=0} \:,
\end{eqnarray}
\label{e:munuc}
\end{mathletters}
whereas the contributions due to $\eps_{NY}$ can be expressed by the 
appropriate components of the single-particle potentials, which 
represent corrections to the Fermi energy of the different species.
This last step amounts to neglecting certain ``rearrangement'' contributions 
that appear in the exact expressions for the chemical potentials, namely
from Eqs.~(\ref{e:epsnny}) one obtains:
\begin{mathletters}
\begin{eqnarray}
  {\partial \eps_{NY} \over \partial \rho_n} &=&
  \sum_{Y=\sim,\la} \left[ U_n^{(Y)}(k_F^{(n)})
  + \sum_{N=n,p} \int_0^{k_F^{(N)}}\!\! {dk\,k^2\over\pi^2} 
  { \partial U_N^{(Y)}(k) \over \partial \rho_n } \right] \:,
\\
  {\partial \eps_{NY} \over \partial \rho_\sim} &=&
  m_\sim + {{k_F^{(\sim)}}^2 \over 2m_\sim} + U_\sim(k_F^{(\sim)}) 
  + \sum_{Y=\sim,\la} \sum_{N=n,p} \int_0^{k_F^{(Y)}}\!\! {dk\,k^2\over\pi^2} 
  { \partial U_Y^{(N)}(k) \over \partial \rho_\sim } \:,
\end{eqnarray}
\end{mathletters}
and similarly for $p$ and $\la$.
Here the last terms represents the rearrangement contributions due 
to the (weak) dependence of $U_N^{(Y)}$ on $\rho_n$
and $U_Y^{(N)}$ on $\rho_\sim$.
In the following, the rearrangement contributions of $\eps_{NY}$ will
be neglected, while the much more important $\eps_{NN}$ is treated
exactly, as specified in Eq.~(\ref{e:munuc}).
This simplifies considerably the numerical effort.

This approximate treatment of $\eps_{NY}$ is justified, apart from the
fact that the nucleon-hyperon forces are weaker than the nucleon-nucleon
forces, by a peculiarity of the BHF approach \cite{bruck},
where the chemical potential $\mu$ of a species is given by 
\begin{equation}
  \mu = e_F + U_2(k_F) + \ldots \:.
\label{e:ef}
\end{equation}
Here $e_F=k_F^2/2m + U(k_F)$ is the Fermi energy, as determined from the 
BHF single-particle potential and $U_2$ is the leading
(of second order in the hole line expansion) rearrangement contribution
to the single-particle potential,
which is given by a diagram representing the conversion of a hole
state into a particle state.
It therefore vanishes in pure neutron matter for all species different 
from the neutron, in particular for the proton and the hyperons.
For the neutron itself, it was shown in Ref.~\cite{nbhf} that 
in pure neutron matter the second order rearrangement contribution
is rather small, due to the relatively weak neutron-neutron interaction.
Therefore, in neutron matter with not too large proton and hyperon fractions,
the hyperon chemical potentials and the corrections to the nucleon
chemical potentials are well approximated by the respective  Fermi energies.

As a further simplification we use the fact that
in the so-called {\it parabolic approximation} \cite{asym}, 
the binding energy per baryon in asymmetric (hyperon-free) nuclear matter 
depends to a good approximation quadratically on the 
asymmetry parameter $\beta=1-2x$: 
\begin{equation}
  \frac{B}{A}(\rho_N,\beta) \approx 
  \frac{B}{A}(\rho_N,\beta=0) + \beta^2 E_{\rm sym}(\rho_N) \:,
\label{e:parab}
\end{equation}
where the symmetry energy $E_{\rm sym}$ can be expressed in
terms of the difference of the energy per particle between pure neutron 
($\beta$=1) and symmetric ($\beta$=0) matter:
\begin{equation}
  E_{\rm sym}(\rho_N) = 
  \frac{B}{A}(\rho_N,\beta=1) - \frac{B}{A}(\rho_N,\beta=0)
  = \frac{1}{2} \frac {\partial(B/A)} {\partial \beta} (\rho_N,\beta=1) \:.
\label{e:sym}
\end{equation}
The composition of neutron star matter is crucially dependent on 
the nuclear symmetry energy. 
This quantity strongly affects the onset of hyperon formation, as well 
as other processes like the neutron star cooling rates \cite{latt}. 
In the parabolic approximation one obtains for the nucleon chemical potentials
\begin{equation}
  \mu_{p,n}(\rho_N,\beta) = \mu_{p,n}(\rho_N,0) 
  -  \left( \beta^2 \pm 2\beta -
  \beta^2 \rho_N \frac{\partial}{\partial\rho_N}  \right) 
   E_{\rm sym}(\rho_N) \:,
\end{equation}
($+$ for $p$, $-$ for $n$), and in particular
\begin{equation}
  [\mu_n-\mu_p](\rho_N,\beta) = 4 \beta E_{\rm sym}(\rho_N) \:.  
\label{eq:mutilda}
\end{equation}

As far as the hyperon chemical potentials, Eq.~(\ref{e:muy}), are concerned,
in practice an effective mass approximation can be employed:
\begin{equation}
  \mu_Y \approx m_Y + U_Y^0 + \frac{(3\pi^2\rho_Y)^{2/3}} {2m_Y^*} \:,
\end{equation}
where the ``mean field'' $U_Y^0 = U_Y(k=0)$ and the global effective mass
\begin{equation}
  \frac{m^\star}{m} = \left[ 1 + {U(k_F) - U(0) \over k_F^2/2m} \right]^{-1}  
\end{equation}
depend on the variables $(\rho,x_p,x_\sim,x_\la)$.
While $U_Y^0$ depends sizeably on all variables,
in practice rather good fits to the calculated single-particle spectra 
were obtained by taking into account only the
$\rho_N$ dependence of $m^*/m$.
Similarly, the corrections to the nucleon Fermi energies appearing in 
Eqs.~(\ref{e:mun},\ref{e:mup}) were found to depend only weakly on 
the proton fraction $x$, and parametrized as function of $\rho_N$ and $x_Y$.

Once the chemical potentials of all species are known, one can proceed
to calculate the composition of stellar matter. 
At high density this is essentially constrained by three conditions: 
i) chemical equilibrium among the different species,
ii) charge neutrality, and iii) baryon number conservation.
The chemical potentials are the fundamental input for solving the equations
for the chemical equilibrium. 
At density $\rho \approx \rho_0$ we assume 
stellar matter composed of a mixture of neutrons, protons,
electrons, and muons in $\beta$-equilibrium [electrons are ultrarelativistic 
at these densities, $\mu_e = (3 \pi^2 \rho x_e)^{1/3}$]. 
In that case the equations read
\begin{eqnarray}
  \mu_n &=& \mu_p + \mu_e \:,
\\
  \mu_e &=& \mu_\mu \:.
\end{eqnarray}
Since we are looking at neutron stars after neutrinos have escaped,
we set the neutrino chemical potential equal to zero. 
Strange baryons appear at density $\rho \approx (2-3) \rho_0$ \cite{bbs}, 
mainly in baryonic
processes like $n + n \rightarrow p + \sim$ and $n + n \rightarrow n + \la$.
The equilibrium conditions for those processes read
\begin{eqnarray}
  2\mu_n &=& \mu_p + \mu_\sim \:,
\\
  \mu_n &=& \mu_\Lambda \:.
\end{eqnarray}
Further two conditions of charge neutrality and baryon number conservation
allow the unique solution of a closed system of equations, 
yielding the equilibrium
fractions of the baryon and lepton species for each fixed baryon density.
They read 
\begin{eqnarray}
   \rho_p  &=& \rho_e + \rho_\mu + \rho_\sim \:,
\label{e:charge}
\\
   \rho &=& \rho_n + \rho_p + \rho_\sim + \rho_\la \:.
\label{e:baryon}
\end{eqnarray}

Finally, from the knowledge of the equilibrium composition one determines
the baryonic equation of state, i.e., the relation between 
baryonic pressure $P_B$ and baryon density $\rho$. 
It can be easily obtained from the thermodynamical relation 
\begin{equation}
  P_B = \rho^2 \frac{d(\epsilon/\rho)}{d\rho} \:.
\end{equation}
The total pressure $P$ and the total mass-energy density $\cal E$
are then calculated by just adding the lepton contributions
that are well-known from textbooks, see e.g. Ref.~\cite{shapiro}:
\begin{eqnarray}
 P &=& P_B + P_l \:,
\\
{\cal E} &=& \eps + \eps_l \:. 
\end{eqnarray}

\section{Results and discussion}
\label{s:res}

\subsection{Single-particle potentials}
\label{s:spp}

\begin{figure}
\includegraphics[totalheight=17.cm,angle=270,bb=180 100 530 800]{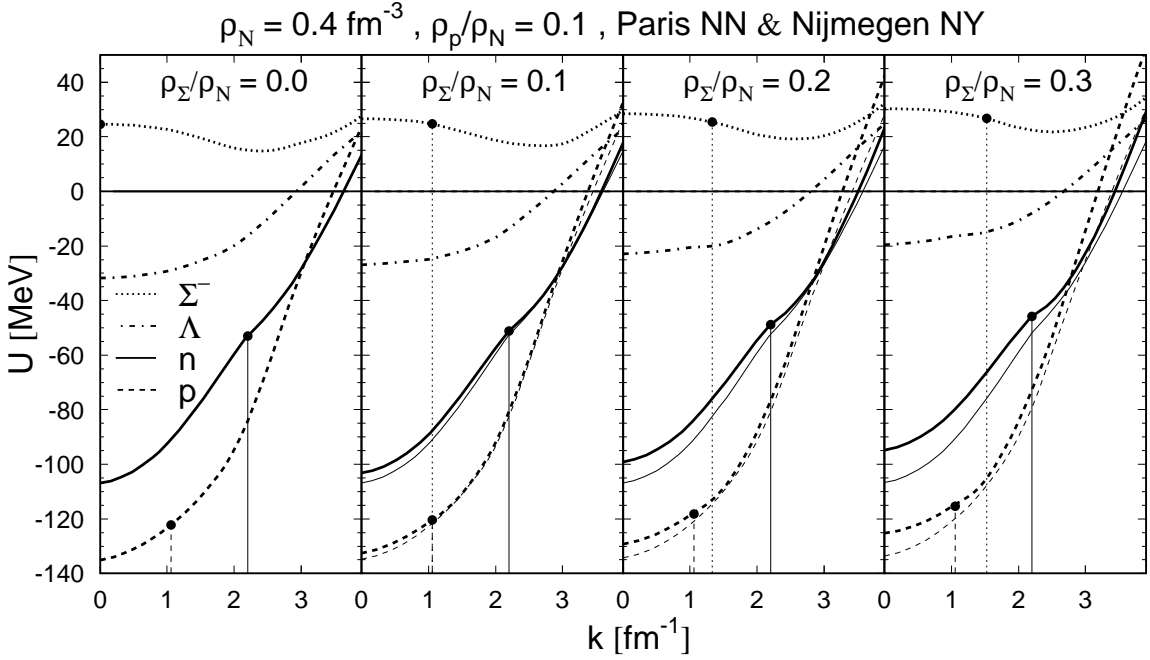}
\caption{
  The single-particle potentials of nucleons $n$, $p$ 
  and hyperons $\sim$, $\la$ in baryonic matter of fixed nucleonic
  density $\rho_N=0.4\,\rm fm^{-3}$, proton density $\rho_p/\rho_N=0.1$,
  and varying $\sim$ density $\rho_\sim/\rho_N=0.0,0.1,0.2,0.3$.
  The vertical lines represent the corresponding Fermi momenta of 
  $n$, $p$, and $\sim$. 
  For the nucleonic curves, the thick lines represent the complete
  single-particle potentials $U_N$, whereas the thin lines show the values 
  excluding the $\sim$ contribution, i.e., $U_N^{(n)} + U_N^{(p)}$.
}
\label{f:u1}
\end{figure}

In order to illustrate some statements made in the previous section, 
we show in Fig.~\ref{f:u1} a representative plot of 
the single-particle potentials of the 
different baryons at fixed neutron and proton densites,
given by $\rho_N=0.4\,\rm fm^{-3}$ and $\rho_p/\rho_N=0.1$,
and varying $\sim$ density.
Under these conditions the  
$\sim$ single-particle potential is sizeably repulsive, while
$U_\la$ is still attractive (see also Ref.~\cite{bbs}) and the nucleons
are much stronger bound.
The $\sim$ single-particle potential has a particular shape with an 
effective mass $m^*/m$ slightly larger than 1,
whereas the lambda effective mass is typically about 0.8 and the
nucleon effective masses are much smaller.

The influence of increasing $\sim$ density on the hyperonic 
single-particle potentials is only indirect 
(since there is no hyperon-hyperon interaction)
and therefore rather small.
There is some additional repulsion for the nucleons
due to the repulsive effective $N\sim$ interaction,
growing with $\sim$ partial density.
However, these effects represent small variations of the 
single-particle potentials observed in the hyperon-free system,
which justifies the approximate treatment  
presented in the previous section.

\begin{figure}
\includegraphics[totalheight=17.cm,angle=270,bb=180 100 530 800]{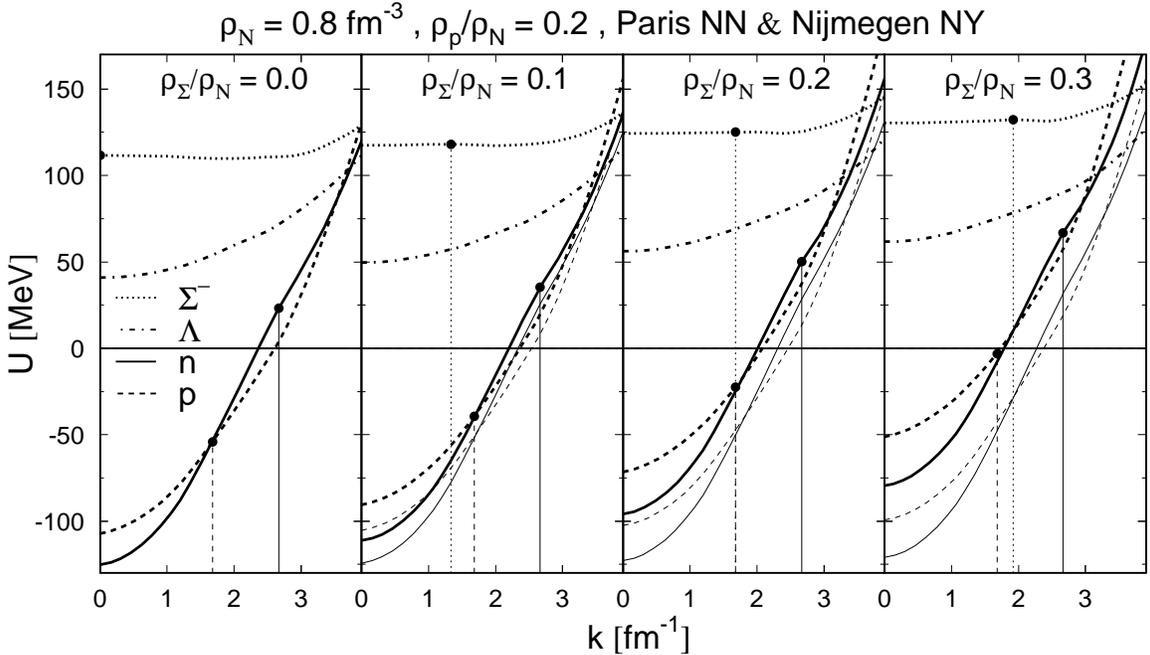}
\caption{
  Same as Fig.~\ref{f:u1}, but for $\rho_N=0.8\,\rm fm^{-3}$ and 
  $\rho_p/\rho_N=0.2$.
}
\label{f:u2}
\end{figure}

The same principal conclusion applies also to Fig.~\ref{f:u2}, that 
displays the same information as Fig.~\ref{f:u1}, but at higher densities
$\rho_N=0.8\,\rm fm^{-3}$ and $\rho_p/\rho_N=0.2$.
The main quantitative difference is the fact that under these conditions
both hyperon single-particle potentials are quite repulsive.

The results displayed were obtained with the Paris nucleon-nucleon potential. 
With the Argonne $v_{18}$ one observes only very slight quantitative changes 
and the corresponding plots are not shown.

\subsection{Stellar matter composition and equation of state}
\label{s:comp}

We come now to the presentation of our results regarding neutron stars. 
In Fig.~\ref{f:1} we show the star composition obtained when the Paris 
(solid line) or the Argonne $v_{18}$ (dotted line) potential 
is adopted as nucleon-nucleon force. 
Three cases are examined, respectively 
(a) no hyperons are present, (b) hyperons are free and (c) hyperons are 
interacting with nucleons. 
In panel (a) we show the particle fractions versus the baryon density 
for matter containing only nucleons and leptons, ignoring hyperons. 
We note that neutrons are the most
dominant species up to very high values of the baryonic density. 
Electrons and muons are present as well. 
These general features are common to many models. 
However, the value of the proton fraction is model dependent, and it 
affects strongly the direct Urca cooling rates \cite{latt}. 

The picture changes when hyperons are taken into account. 
This is shown in panels (b) and (c). 
The $\sim$ is the first hyperon to appear, due to its negative
charge, whereas $\la$ formation takes place at higher density.
Other hyperon species do not appear in our model.
In the free hyperon case [panel (b)], the formation of $\sim$ starts at about
$\rho \approx 0.4 \,\rm fm^{-3}$, as was published in our previous 
paper \cite{bbs}.
In that paper the $\la$ onset point was approximate, because for its 
precise determination all the chemical potentials at finite $\sim$ fraction 
are needed. 
This is performed in the present calculations, and now 
we estimate exactly the $\la$ onset point, which is located at about 
$\rho \approx 0.82\,\rm fm^{-3}$. 
As is clearly shown, the thresholds for hyperon
formation are weakly dependent on the two-body force. 
The hyperon fractions are substantial
at high density and constitute a large portion of the stellar core matter. 
The appearance of hyperons induces a deleptonization of the
baryonic medium, mainly because of the charge neutrality condition.
Leptons disappear at high baryonic density, thus hindering formation of 
kaon condensate \cite{kaon}. 

When the nucleon-hyperon interaction is taken into account,
the scenario described above changes quantitatively. 
In fact, since the $\sim$ Fermi energy is 
repulsive starting from densities just above normal nuclear matter
density \cite{bbs}, the onset point is shifted to slightly higher density. 
On the contrary, the $\la$ formation starts at density lower than in the 
free case because the $\la$ Fermi energy is attractive over a wide
range of densities.
This is clearly shown in panel (c). 
However, the hyperon population is smaller than in the free case and 
deleptonization is less drastic, because the repulsive core of the NY
interactions becomes relevant at high density. 
However, even in this case, kaon condensation cannot occur. 
Again, we note how those results are 
slightly dependent on the two-body interaction. 

The main consequence of introducing additional particle species
into matter is the softening of the EOS. 
This softening is essentially
due to the conversion of kinetic energy of the already
present species into masses of the new species. 
The decrease of the lepton population does add further softening, but 
this effect is quite small. 
The nature of the EOS is thus dependent on the
number of species, as well as on the details of the strong interaction. 
It can be instructive to begin with the equation of state corresponding
to the case (a), i.e., when no hyperons are present. 
This is displayed in Fig.~\ref{f:2} (solid line). 
On the left (right) hand side we show the equation 
of state obtained when the Paris ($Av_{18}$) potential is adopted
as nucleon-nucleon interaction.
That calculation has been improved with 
respect to the one published in Ref.~\cite{threeb}, since more channels
are now taken into account in the solution of the Bethe-Goldstone
equation.
This has produced a better convergence of the iterative procedure for
high values of nuclear matter density. 

The properties of the EOS change when hyperons are taken into account. 
In particular, the dotted line corresponds to the case of nuclear
matter containing leptons and free hyperons. 
The presence of hyperons induces a strong softening of the equation of state.
The inclusion of the nucleon-hyperon interaction produces an
equation of state stiffer than in the free hyperon case. 
This is shown by the dashed line. 
We observe a similar behavior when the Argonne $v_{18}$ potential is used. 

However, it is well known that 
nonrelativistic calculations, based on purely two-body interactions, fail 
to reproduce the correct saturation point of symmetric nuclear matter
\cite{coester}. 
This deficiency is commonly corrected by
introducing three-body forces (TBF) among nucleons. 
This changes the scenario described above. 
Our method of treating TBF is discussed in the following paragraph.

\subsection{Inclusion of three-body forces}
\label{s:tbf}

It is commonly known that a complete theory of three-body forces is not 
available so far. 
Therefore one has to work with phenomenological approaches. 
A realistic model for nuclear TBF is the so-called Urbana model
\cite{schi}, which consists of an attractive term due to two-pion exchange
with excitation of an intermediate $\Delta$ resonance, and a repulsive 
phenomenological central term. 
We introduced the same Urbana three-nucleon
model within the BHF approach (for more details see Ref.~\cite{threeb}).
In our approach the TBF is reduced to a density dependent two-body force by
averaging on the position of the third particle, assuming that the
probability of having two particles at a given distance is reduced 
according to the two-body correlation function. 
The corresponding EOS satisfies several requirements, namely
(i) it reproduces correctly the nuclear matter saturation point \cite{threeb},
(ii) the incompressibility  is compatible
with values extracted from phenomenology \cite{myers}, 
(iii) the symmetry energy is compatible with nuclear phenomenology, 
(iv) the causality condition is always fulfilled.  

Fig.~\ref{f:3} shows the values of the symmetry energy for the 
different EOS's that we consider, namely 
the nonrelativistic Brueckner calculations with the Paris and the
Argonne $v_{18}$ potentials with and without three-body forces.
For comparison, we report also the symmetry energy 
of a recent calculation performed with a Dirac-Brueckner (DBHF)
model \cite{dbhf}, but with the Bonn-A potential.
We should remind that the DBHF treatment is equivalent \cite{zdia} to 
introducing in the nonrelativistic BHF the three-body force corresponding to 
the excitation of a nucleon-antinucleon pair, 
the so-called Z-diagram \cite{brown}, which is repulsive at all densities. 
We therefore expect that the symmetry energy calculated within the DBHF
approach is always larger than the one obtained in the nonrelativistic
BHF calculation, without and with three-body force, since in the
last case both an attractive and a repulsive component is introduced.
From Fig.~\ref{f:3} we see that this is true for all the potentials
discussed in this paper, besides for the Argonne $v_{18}$ potential with TBF,
maybe because of its strongly repulsive core. 
On the contrary, in the low density region 
($\rho \lesssim 0.3\;{\rm fm}^{-3}$), 
both BHF+TBF symmetry energies and DBHF calculations are very similar.

We can proceed now to the discussion of the nuclear matter composition
when TBF are included in the equation of state. 
Please note that TBF are included only for nucleons. 
Hyperons interact via two-body forces with nucleons 
(we adopt the Nijmegen soft-core potential, as discussed in the introduction),
and do not interact at all among themselves. 
Since no experimentally tested hyperon-hyperon potentials
are currently available, 
this assumption is in line with the exploratory character of this work.
Of course, their eventual introduction may change the scenarios 
resulting from our analysis.

\begin{figure}
\includegraphics[totalheight=19.5cm,angle=270,bb=150 50 500 750]{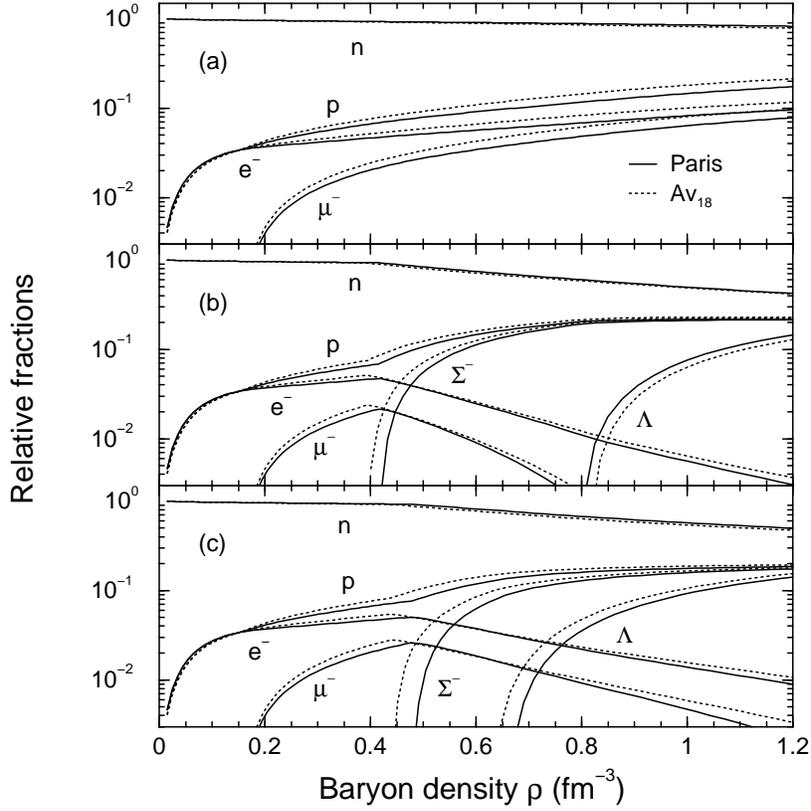}
\caption{
 Composition of neutron star matter in the BHF model.
 The top panel (a) shows the results without hyperons; 
 the middle panel (b) results with noninteracting hyperons; 
 and the lower panel (c) includes interaction between nucleons and hyperons.
}
\label{f:1}
\end{figure}

\begin{figure}
\includegraphics[totalheight=19.5cm,angle=270,bb=180 100 480 800]{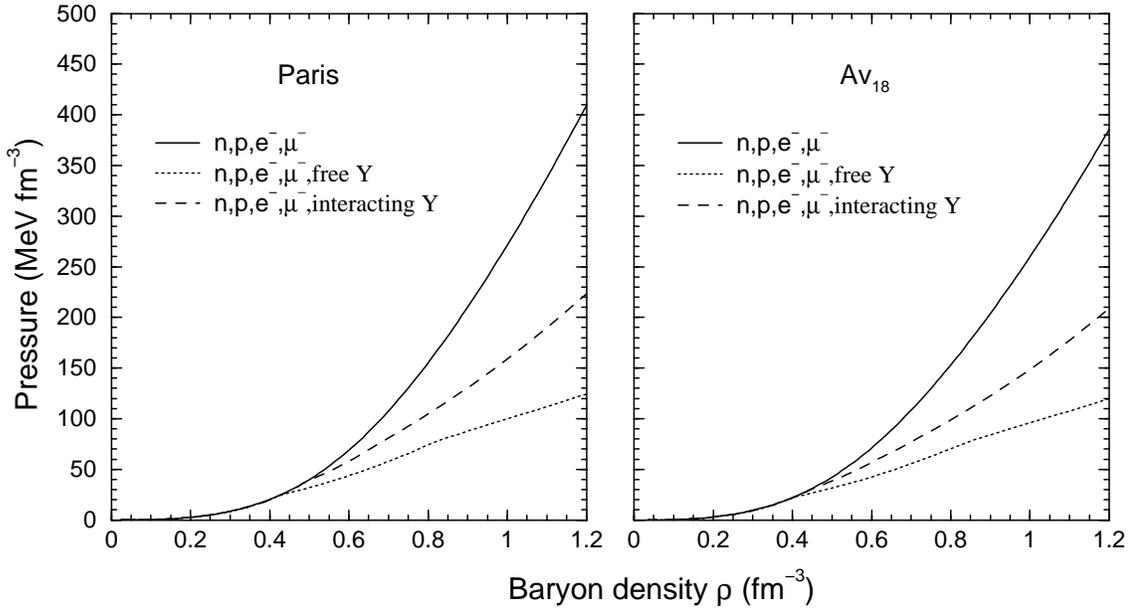}
\caption{
 EOS of neutron star matter in the BHF model.
 Results without hyperons (solid lines), with free hyperons (dotted lines),
 and with interacting hyperons (dashed lines) are shown.
 The Paris (left) or Argonne $v_{18}$ (right) nucleon-nucleon potentials 
 were used.
}
\label{f:2}
\end{figure}

\begin{figure}
\includegraphics[totalheight=19.cm,angle=270,bb=180 150 530 850]{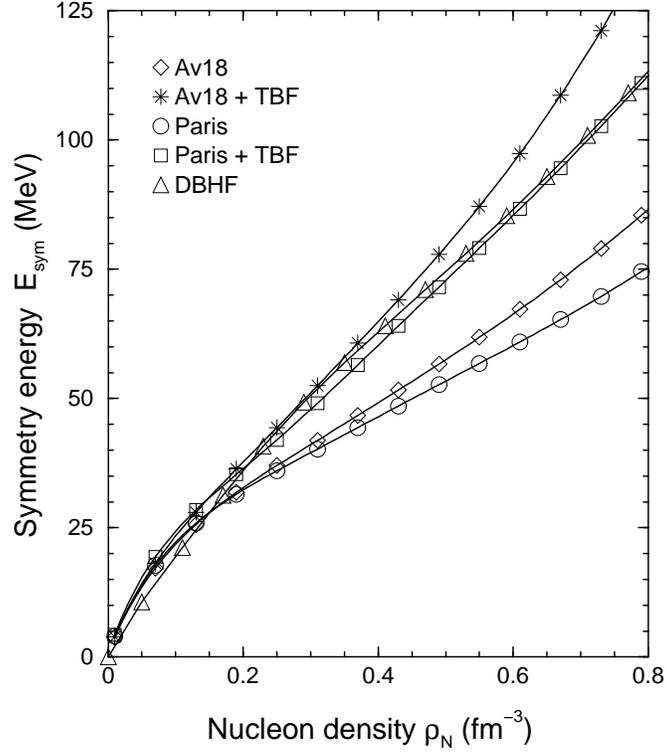}
\caption{
 Nucleonic symmetry energy as a function of nucleon density
 obtained in different theoretical models of nuclear matter.
}
\label{f:3}
\end{figure}

\begin{figure}
\includegraphics[totalheight=19.cm,angle=270,bb=120 50 500 750]{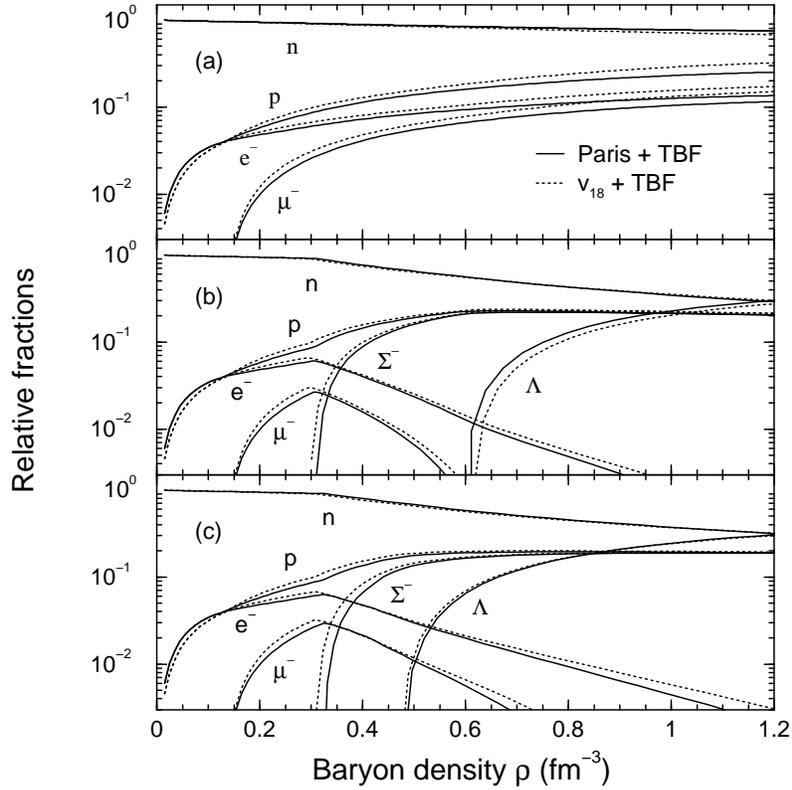}
\caption{
 As Fig.~\ref{f:1}, but the nucleon-nucleon potentials are supplemented
 by three-body forces.
}
\label{f:4}
\end{figure}

In Fig.~\ref{f:4} we show the matter composition when the TBF 
among nucleons is included in the BHF calculation. 
The notation is the same as in Fig.~\ref{f:1}, i.e., solid line 
represents the composition obtained
with the Paris+TBF potential, whereas the dotted line corresponds to the
composition obtained with the Argonne $Av_{18}$ potential.
Let us first discuss the case without hyperons, displayed in panel (a). 
Because of its repulsive
character at high densities, the higher value of the symmetry energy
allows more easily the conversion of neutrons into protons and leptons
compared to the case without TBF. 
The proton fraction can now 
exceed the ``critical'' value $x_{\rm Urca} \approx (11-15) \%$ 
needed for the occurrence of direct Urca processes \cite{latt}. 

The chemical composition of nuclear matter changes completely 
when hyperons are taken into account. 
In panel (b) we show the particle fractions obtained when the 
hyperons are free. 
We notice that the three-body forces shift the onset points of both 
$\sim$ and $\la$, as previously published \cite{bbs}, down to 
densities 2--3.5 times normal nuclear matter density. 
Deleptonization takes place, and leptons disappear
almost completely just after hyperon formation, because now it is 
more convenient energetically to maintain the charge neutrality through 
$\sim$ formation than $\beta$-decay. 
At high density nucleons and hyperons are present almost in the
same percentage.
 
This scenario does not change qualitatively when the 
nucleon-hyperon interaction is included, see panel (c). 
The main difference is that the hyperon onset points move again, 
respectively to higher (lower)
density because of the repulsive (attractive) character of the $\sim$ ($\la$)
Fermi energy at those densities. 
Even in this case the stellar core is composed
by an almost equal fraction of nucleons and hyperons.
Once again those compositions look very much the same both with the Paris 
and the $Av_{18}$ potential.

\begin{figure}
\includegraphics[totalheight=19.5cm,angle=270,bb=180 100 480 800]{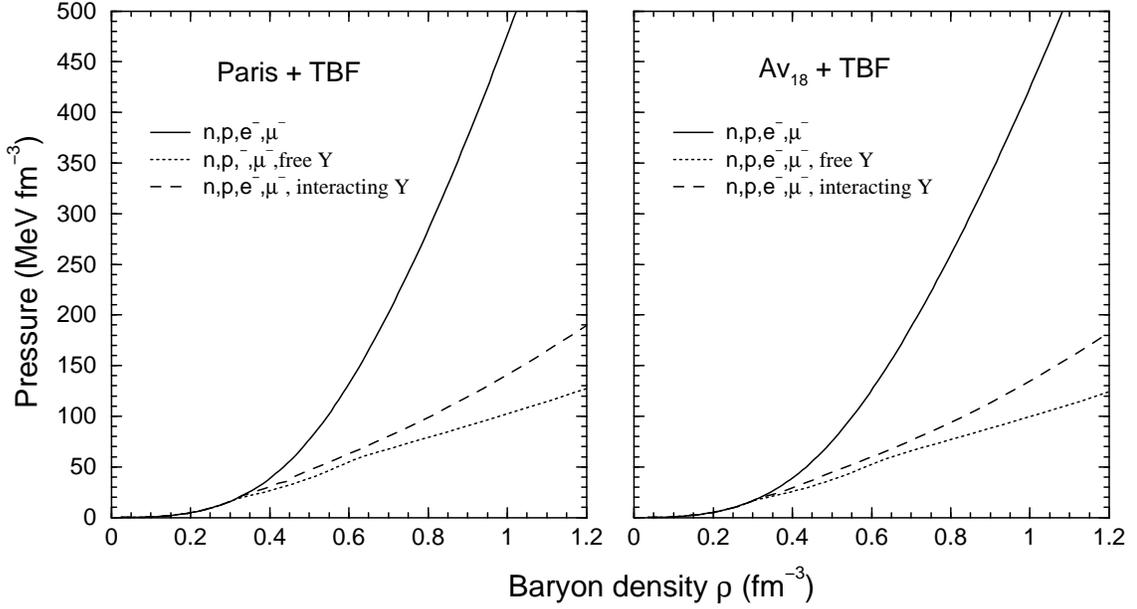}
\caption{
 As Fig.~\ref{f:2}, but including three-body forces.
}
\label{f:5}
\end{figure}

The corresponding equations of state obtained when TBF are added to the 
two-body forces are listed in Table~\ref{t:1} and shown in Fig.~\ref{f:5}. 
On the left (right) hand side we show 
our results obtained with the Paris ($Av_{18}$) potential plus TBF.
The solid line shows the equation of state of asymmetric beta-stable matter
with a percentage of electrons and muons. 
It looks much stiffer than the case when hyperons are introduced. 
The dotted line represents the equation of 
state of nuclear matter containing free hyperons, whereas the dashed line
corresponds to the case of hyperons interacting with the nucleonic medium.
If we compare the equations of state obtained without and
with TBF, i.e., the dashed lines of Figs.~\ref{f:2} and \ref{f:5}, 
we see only a small difference. 
This can be understood by looking at the final compositions,
respectively panels (c) of Figs.~\ref{f:1} and \ref{f:4}. 
There we note that the net effect of TBF is a
decreased presence of neutrons and an enhanced population of $\la$, whereas 
protons and $\sim$ are only slightly affected by TBF, keeping the 
total percentage of neutral charge  almost the same in both cases.
Therefore, as far as the equation of state is concerned, we expect 
that the relation between pressure and baryon density is very similar 
without and with TBF. 
This is indeed the case and means that hyperon formation is a mechanism 
for pressure control, as already found by 
the authors of Ref.~\cite{pheno} within a phenomenological approach.

\begin{figure}
\includegraphics[totalheight=17.5cm,angle=270,bb=80 80 480 750]{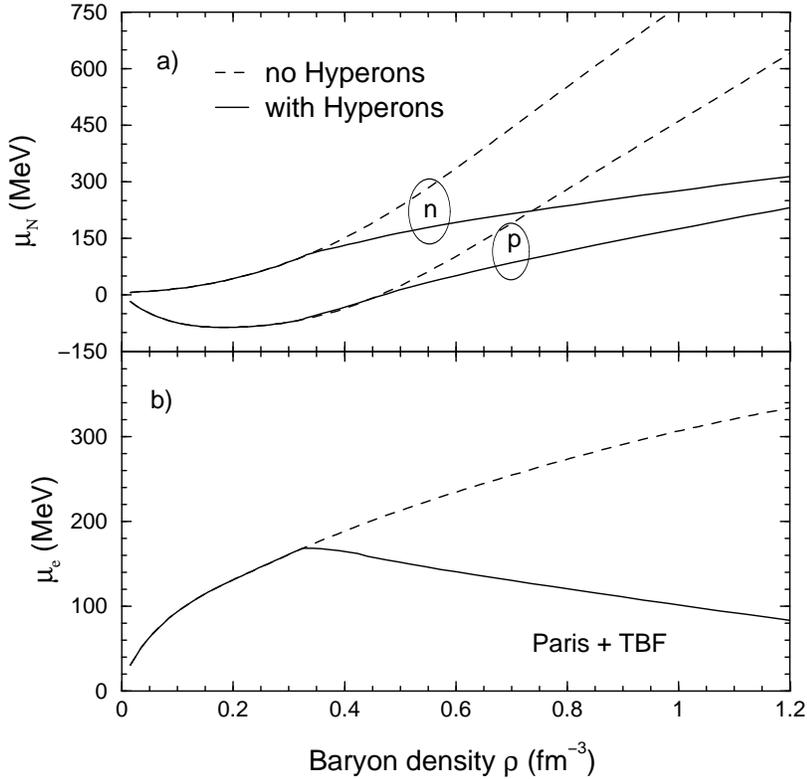}
\caption{
 Chemical potentials of the different species as a function of baryon
 density in neutron star matter. 
 The upper panel (a) shows results for neutrons and protons, whereas the 
 electron chemical potential is displayed in panel (b). 
 The dashed (solid) line denotes the case of matter 
 composition without (with) interacting hyperons.}
\label{f:mu}
\end{figure}

Finally, we comment on the deleptonization of the baryonic matter. 
This is clearly indicated by the chemical potentials, drawn in Fig.~\ref{f:mu}.
In panel (a) we display the chemical potentials for neutrons and protons
obtained with the Paris potential supplemented by TBF (the value of the
neutron mass has been subtracted), whereas in panel (b) the electron chemical 
potential is shown. 
We observe a monotonously increasing function of the baryon density when 
no hyperons are present (dashed lines), whereas 
the appearance of negatively charged hyperons interacting
with the medium (solid line) induces a strong deleptonization of the matter.
Thus the formation of hyperons and the consequent deleptonization will 
suppress meson condensation \cite{ellis,mishu}. 
Although kaon and anti-kaon condensation has been
proposed as possible state of matter inside neutron stars \cite{kaon},
also recent investigations 
within the relativistic mean field approach \cite{mishu}  
find the onset of kaon condensation quite unlikely.
In any case, the importance of meson condensation will be strongly
diminished by allowing for the dominant hyperon formation.

\subsection{Neutron stars}
\label{s:nstar}

As already discussed in the introduction, the knowledge of the equation 
of state is essential in order to build models of both static and 
rotating neutron stars. 
In fact, the EOS for $\beta$-stable matter can be used in the 
Tolman-Oppenheimer-Volkoff (TOV) equations \cite{tol,ov} 
to compute the neutron star 
mass and radius as a function of the central density. 
As a first step, we neglect the effects of rotations and calculate
the mass-radius relation assuming that a neutron star is a spherically 
symmetric object in hydrostatic equilibrium. 
Then the equilibrium configurations are simply obtained by solving the TOV 
equations for the total pressure $P$ and the enclosed mass $m$, 
\begin{eqnarray}
  {dP(r)\over{dr}} &=& -{ G m(r) {\cal E}(r) \over r^2 } \,
  {  \left[ 1 + \frac{P(r)}{{\cal E}(r)} \right] 
  \left[ 1 + \frac{4\pi r^3 P(r)}{m(r)} \right] 
  \over
  1 - \frac{2 G m(r)}{r} } \:,
\\
  {dm(r) \over dr} &=& 4 \pi r^2 {\cal E}(r) \:,
\end{eqnarray}
being $G$ the gravitational constant (we assume $c=1$). 
Starting with a central mass density ${\cal E}(r=0) \equiv {\cal E}_c$,  
we integrate out until the pressure on the surface equals the one 
corresponding to the density of iron.
This gives the stellar radius $R$ and the gravitational mass is then 
\begin{equation}
M_G~ \equiv ~ m(R)  = 4\pi \int_0^R dr\,r^2 {\cal E}(r) \:. 
\end{equation}
For the outer part of the neutron star we have used the equations of state
by Feynman-Metropolis-Teller \cite{fey} and Baym-Pethick-Sutherland 
\cite{baym}, 
and for the middle-density regime 
($0.001\,{\rm fm^{-3}} < \rho < 0.08\,{\rm fm}^{-3}$) 
we use the results of Negele and Vautherin \cite{nv}. 
In the high-density part 
($\rho > 0.08\,{\rm fm}^{-3}$) 
we use alternatively the equations of state displayed in Table~\ref{t:1}. 
For comparison, we also perform calculations of neutron star structure 
for the case of purely nucleonic asymmetric and $\beta$-stable matter
with some lepton fraction.   

\begin{figure}
\includegraphics[totalheight=11.5cm,angle=90,bb=50 150 575 700]{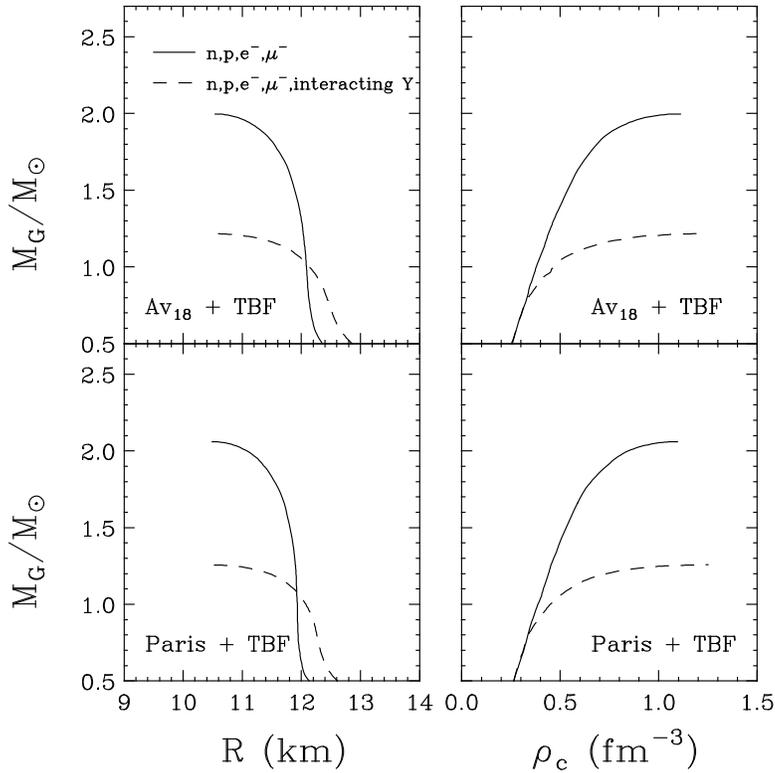}
\caption{
 Neutron star mass as a function of radius (left) or central baryon
 density (right) in the BHF+TBF model.
 Results without (solid lines) and with interacting hyperons (dashed lines)
 are compared. 
}
\label{f:6}
\end{figure}

The results are plotted in Fig.~\ref{f:6}, 
where we display the gravitational mass $M_G$ 
(in units of the solar mass $M_\odot$)
as a function of the radius $R$ and the central baryon density 
$\rho_c$. 
We report the results obtained using either the $Av_{18}$ 
or the Paris two-body potential with three-body forces.
The solid lines represent the equilibrium configurations
of neutron stars composed only of nucleons and leptons; the dashed lines 
show the configurations of stars whose composition includes interacting 
hyperons. 
As we can see, the softening of the equation of state due to the
presence of additional baryonic species produces a strong decrease in the value
of the limiting mass, from about $2 M_\odot$ down to $1.2 M_\odot$,
almost independent on the nucleon-nucleon interaction. 
The limiting central densities stay nearly constant at about 7 times normal 
nuclear matter density.

For clarity we report in Table~\ref{t:2} the properties of the maximum mass
configuration obtained with equations of state of $\beta$-stable 
asymmetric nuclear matter with and without hyperons. 
Without hyperons we observe mainly 
the effect of three-body forces among nucleons on the equation of state and,
therefore, on the values of the maximum mass configuration. 
In particular, the increased repulsion among nucleons produces a stiffer 
equation of state and a higher value of the maximum mass, 
with a smaller radius and central density. 
Even in this case we note a strong similarity in the limiting values
independently on the two-body potential. 

The situation changes dramatically when interacting hyperons are included. 
In this case the additional repulsion produced by
the nucleon three-body forces is counterbalanced by the increased 
population of hyperons, thus leading to a very soft equation of state. 
The values of the maximum mass do not differ very much 
from those calculated without TBF. 
However, the values of the maximum central densities are substantially 
different, despite the EOS looking similar in the two cases. 
When TBF are included, this value becomes smaller, see Table~\ref{t:2}. 
Without TBF, in fact, the EOS is slightly stiffer at higher densities
and the corresponding curve of the mass vs.~central density 
(not reported in Fig. 9) has a shallow maximum at larger central density. 
However, the corresponding maximum mass is only slightly
larger than in the case with TBF, as reported in Table~\ref{t:2}.
This also shows the sensitivity of the results on the details of the
nuclear EOS. 
It has to be noticed that, for stable configurations, at a given value of
the neutron star mass the central density is larger when TBF are included.
Furthermore, neutron stars built with equations of state of matter containing
baryons interacting via TBF possess a core with a larger hyperon population
than in the case without TBF.

\begin{figure}
\includegraphics[totalheight=12.5cm,angle=90,bb=60 110 410 680]{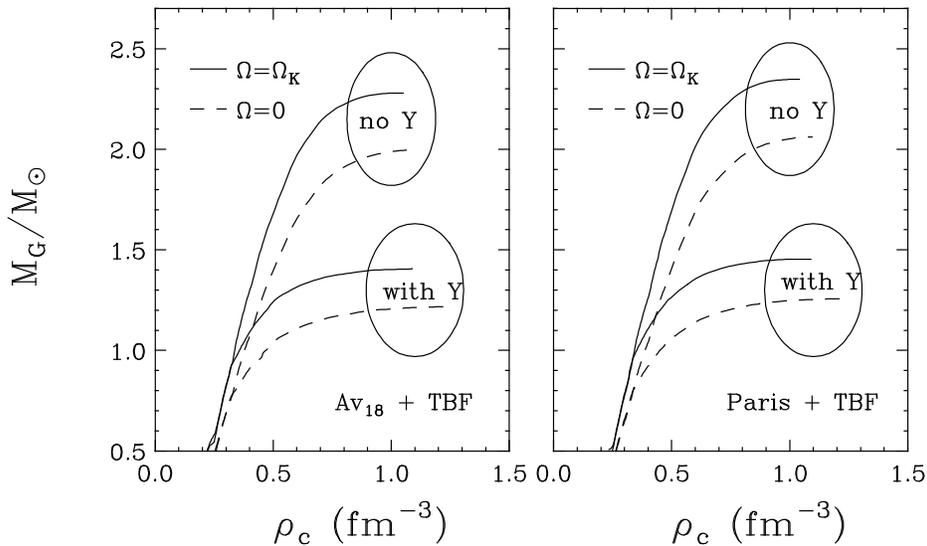}
\caption{
 Neutron star mass as a function of central baryon density in the BHF+TBF
 model with and without hyperons.
 Curves for nonrotating stars (dashed lines) and rotations at the 
 Kepler frequency (solid lines) are shown.
}
\label{f:7}
\end{figure}

The above scenario changes when the rotations are included. 
In particular, in order to treat stellar rotations in general relativity, 
we follow the method discussed in Ref.~\cite{wegl}.
In those papers the Einstein equations for rotating massive objects are
solved perturbatively with a modified version of Hartle's method. 
With this method the authors are able to determine sequences of star 
models rotating at their respective general relativistic 
Kepler frequency $\Omega_K$. 
In Fig.~\ref{f:7} we display the net effect of 
rotations on the mass-density relation. 
There we display star sequences at equilibrium both for static (dashed lines) 
and rotating (solid lines) neutron stars. 
The upper curves 
refer to equations of state without hyperons, whereas the lower curves  
refer to equations of state containing hyperons. 
The rotations increase the value of the limiting mass by about $12 \%$, 
decreasing the limiting value of the central density.
Again, we do not observe a significant difference for the various 
NN potentials.
In Table~\ref{t:2} we report also the values of the limiting configuration.

\section{Conclusions}
\label{s:conc}

The principal finding of this paper
has been a suprisingly low value for the maximum mass of a ``neutron star''
that barely comprises the ``canonical'' value $M \approx 1.4 M_\odot$.
While certainly the technical insufficiences and approximations of 
our approach can account for an uncertainty on this limit of a few percent,
it seems rather difficult theoretically to avoid a rather low limit
on the maximum mass. 
As we have seen, the mere presence of additional baryonic degrees of freedom
in the form of hyperons renders the maximum mass quite insensitive to 
the stiffness of the {\em nucleonic} equation of state.
Because the hyperonic onset densities at about 2--3 times normal density
seem to be rather robust and model independent, it seems that the
maximum mass can be substantially increased only if the nucleon-hyperon
and/or hyperon-hyperon effective interactions become extremely repulsive 
at high density.
Unfortunately at present the experimental information on the 
nucleon-hyperon interaction is rather scarce, and for the hyperon-hyperon
case practically non-existent.

Nevertheless, recently the Nijmegen NY potential model was extended to the
full YY case by imposing isospin symmetry \cite{stoks}.
In Ref.~\cite{barc} these new potentials have been used 
to include the hyperon-hyperon interaction in neutron star studies
within a similar approach,
using a variational nucleonic EOS together with hyperon single-particle
potentials determined in a standard choice Brueckner scheme.
The obtained maximum neutron star masses turn out to be quite similar to 
the ones presented in this work.

Let us also mention that, within different versions of the relativistic
mean field theory, the introduction of hyperons in the EOS of
nuclear matter leads to maximum masses that are only slightly larger
\cite{greiner} or substantially larger \cite{njl} than the ones reported here. 
This illustrates the lack of constraints imposed on this type of models.

In Ref.~\cite{njl} the possible onset of quark-gluon plasma in the 
interior of neutron stars
is also investigated and found to be unlikely to occur up to 
baryon densities we found in the interior of the neutron star. 
It is expected that the possible presence
of the deconfined phase would produce an additional softening  
of the EOS, and therefore a further decrease of the critical mass. 

It seems, therefore, that indeed the theoretical limit of the 
maximum mass of a neutron star is rather small.
A lowering of the maximum mass of neutron stars due to the onset of 
hyperon formation was actually already qualitatively
anticipated in the very first articles which investigated 
baryonic matter with hyperons \cite{nstar}. 
However, no quantitative analysis could be done at that time, due to
the uncertainties in the theoretical models.
Within the present microscopic approach
we can give now a more quantitative prediction of this effect.

\section{Acknowledgements}
We warmly thank F. Weber, who provided us with his code for the
calculation of the structure of rotating neutron stars 
and acknowledge useful discussions with A. Polls and A. Ramos.
This work was supported in part by the program
``Estancias de cient\'ificos y tecn\'ologos extranjeros en Espa\~na''.

\bigskip
\bigskip
\bigskip





\begin{table}
\caption{
EOS for $\beta$--stable matter with hyperons obtained in the BHF 
approximation using the Argonne $v_{18}$ or Paris two-body interaction 
complemented by the Urbana model for three-body forces. 
We display the baryon density $\rho$, the total mass density ${\cal E}$, 
and the total pressure $P$.}
\bigskip
\begin{tabular}{|d|dd|dd|}
$\rho\ (\rm fm^{-3})$ & 
\multicolumn{2}{c|}{${\cal E}~(10^{14}\,\rm g\,cm^{-3})$} & 
\multicolumn{2}{c|}{$P~(10^{34}\,\rm dyn\,cm^{-2})$} \\
 & $Av_{18}$+TBF & Paris+TBF & $Av_{18}$+TBF & Paris+TBF \\
\hline
0.08  &  1.27 & 1.27 &   0.07 & 0.06  \\
0.16  &  2.64 & 2.63 &   0.42 & 0.39  \\
0.2   &  3.33 & 3.33 &   0.80 & 0.75  \\
0.3   &  5.13 & 5.12 &   2.62 & 2.52  \\
0.4   &  7.11 & 7.20 &   4.70 & 5.05  \\
0.5   &  9.01 & 8.96 &   7.16 & 7.40  \\
0.6   &  11.1 & 11.0 &   9.71 & 10.1  \\
0.7   &  13.0 & 13.1 &   12.2 & 12.8  \\
0.8   &  15.3 & 15.2 &   15.3 & 16.2  \\
0.9   &  17.3 & 17.2 &   18.3 & 18.9  \\
1.0   &  20.0 & 20.1 &   22.5 & 23.5  \\
1.1   &  21.8 & 22.1 &   25.7 & 27.2  \\
1.2   &  23.9 & 24.4 &   29.3 & 30.5  \\
1.3   &  26.0 & 26.2 &   33.3 & 34.5  \\
1.4   &  28.8 & 28.5 &   38.8 & 39.2  \\
1.5   &  31.1 & 31.1 &   43.7 & 45.2  \\
1.6   &  33.5 & 33.4 &   49.0 & 49.7  \\
1.7   &  35.4 & 36.2 &   53.2 & 56.2  \\
\end{tabular}
\label{t:1}
\end{table}

\begin{table}
\caption{
Properties of the maximum mass configuration obtained for different 
equations of state: $M_G$ is the gravitational (maximum) mass,
$R$ is the corresponding radius, and $\rho_c$ the central 
baryon density.
In each case the results of the EOS without hyperons (no Y), including
hyperons (Y), and including rotation at the Kepler frequency (Y+Rot)
are listed.}
\bigskip
\begin{tabular}{|c|ddd|ddd|ddd|}
EOS & 
\multicolumn{3}{c|}{$M_G/M_\odot$} & 
\multicolumn{3}{c|}{$R$ (km)} &  
\multicolumn{3}{c|}{$\rho_c$ (fm$^{-3}$)} \\
 & no Y & Y & {Y+Rot} & no Y & Y & {Y+Rot} & no Y & Y & {Y+Rot} \\
\hline
 $Av_{18}$     &1.64 &1.26 &1.44  & 9.10 & 8.70 &10.53 &1.53 &1.86 &1.56\\
 Paris         &1.67 &1.31 &1.50  & 8.90 & 8.62 &10.30 &1.59 &1.84 &1.62 \\
 $Av_{18}$+TBF &2.00 &1.22 &1.41  &10.54 &10.46 &12.68 &1.11 &1.25 &1.09 \\
 Paris+TBF     &2.06 &1.26 &1.45  &10.50 &10.46 &12.65 &1.10 &1.25 &1.09 \\
\end{tabular}
\label{t:2}
\end{table}

\end{document}